\documentstyle[12pt,psfig]{article}
\def\simlt{\stackrel{<}{{}_\sim}}
\def\simgt{\stackrel{>}{{}_\sim}}
\def\be{\begin{equation}}
\def\ee{\end{equation}}
\def\bea{\begin{eqnarray}}
\def\eea{\end{eqnarray}}
\def\simlt{\stackrel{<}{{}_\sim}}
\def\simgt{\stackrel{>}{{}_\sim}}
\def\ap#1#2#3   {{\em Ann. Phys. (NY)} {\bf#1} (#2) #3.}
\def\apj#1#2#3  {{\em  Astrophys. J.} {\bf#1} (#2) #3.}
\def\apjl#1#2#3 {{\em Astrophys. J. Lett.} {\bf#1} (#2) #3.}
\def\app#1#2#3  {{\em Acta. Phys. Pol.} {\bf#1} (#2) #3.}
\def\ar#1#2#3   {{\em Ann. Rev. Nucl. Part. Sci.} {\bf#1} (#2) #3.}
\def\cpc#1#2#3  {{\em Computer Phys. Comm.} {\bf#1} (#2) #3.}
\def\err#1#2#3  {{\em Erratum} {\bf#1} (#2) #3.}
\def\ib#1#2#3   {{\em ibid.} {\bf#1} (#2) #3.}
\def\jmp#1#2#3  {{\em J. Math. Phys.} {\bf#1} (#2) #3.}
\def\ijmp#1#2#3 {{\em Int. J. Mod. Phys.} {\bf#1} (#2) #3.}
\def\jetp#1#2#3 {{\em JETP Lett.} {\bf#1} (#2) #3.}
\def\jpg#1#2#3  {{\em J. Phys. G.} {\bf#1} (#2) #3.}
\def\mpl#1#2#3  {{\em Mod. Phys. Lett.} {\bf#1} (#2) #3.}
\def\nat#1#2#3  {{\em Nature (London)} {\bf#1} (#2) #3.}
\def\nc#1#2#3   {{\em Nuovo Cim.} {\bf#1} (#2) #3.}
\def\nim#1#2#3  {{\em Nucl. Instr. Meth.} {\bf#1} (#2) #3.}
\def\np#1#2#3   {{\em Nucl. Phys.} {\bf#1} (#2) #3.}
\def\pcps#1#2#3 {{\em Proc. Cam. Phil. Soc.} {\bf#1} (#2) #3.}
\def\pl#1#2#3   {{\em Phys. Lett.} {\bf#1} (#2) #3.}
\def\prep#1#2#3 {{\em Phys. Rep.} {\bf#1} (#2) #3.}
\def\prev#1#2#3 {{\em Phys. Rev.} {\bf#1} (#2) #3.}
\def\prl#1#2#3  {{\em Phys. Rev. Lett.} {\bf#1} (#2) #3.}
\def\prs#1#2#3  {{\em Proc. Roy. Soc.} {\bf#1} (#2) #3.}
\def\ptp#1#2#3  {{\em Prog. Th. Phys.} {\bf#1} (#2) #3.}
\def\ps#1#2#3   {{\em Physica Scripta} {\bf#1} (#2) #3.}
\def\rmp#1#2#3  {{\em Rev. Mod. Phys.} {\bf#1} (#2) #3.}
\def\rpp#1#2#3  {{\em Rep. Prog. Phys.} {\bf#1} (#2) #3.}
\def\sjnp#1#2#3 {{\em Sov. J. Nucl. Phys.} {\bf#1} (#2) #3.}
\def\spj#1#2#3  {{\em Sov. Phys. JEPT} {\bf#1} (#2) #3.}
\def\spu#1#2#3  {{\em Sov. Phys.-Usp.} {\bf#1} (#2) #3.}
\def\zp#1#2#3   {{\em Zeit. Phys.} {\bf#1} (#2) #3.}

\def\NPB#1#2#3{{\it Nucl.~Phys.} {\bf{B#1}} (19#2) #3}
\def\PLB#1#2#3{{\it Phys.~Lett.} {\bf{B#1}} (19#2) #3}
\def\PRD#1#2#3{{\it Phys.~Rev.} {\bf{D#1}} (19#2) #3}
\def\PRL#1#2#3{{\it Phys.~Rev.~Lett.} {\bf{#1}} (19#2) #3}

\begin{document}
\topmargin-2.5cm
\begin{titlepage}
\begin{flushright}
CERN-TH/95-319\\
\end{flushright}
\vskip 0.3in
\begin{center}{\Large\bf 
UNIFICATION OF COUPLINGS IN THE MSSM} \footnote{To appear 
in the Proceedings of the Brussels EPS-HEP 95 Conference, Aug.-Sep. 1995,
Brussels, Belgium.} \\
\vskip .5in
{\bf M. Carena} \vskip.35in
CERN, TH Division, CH--1211 Geneva 23, Switzerland\\
\end{center}
\vskip2.cm
\begin{center}
{\bf Abstract}
\end{center}
\begin{quote}
 I briefly summarize  the main features
related to the unification of gauge and Yukawa couplings in the MSSM,
emphasizing  the predictions derived in this framework, which are of special
interest in the light of present and near--future experiments
\end{quote}
\vskip1.cm

\end{titlepage}
\setcounter{footnote}{0}
\setcounter{page}{1}
\newpage

The Minimal Supersymmetric extension
of the Standard Model (MSSM)  provides a solution to the  hierarchy
problem of the Standard Model (SM),  related to
the existence of quadratic divergences  associated with the vacuum
expectation value of the Higgs field which  determines
 the fermion and gauge boson masses.
Moreover, the MSSM can be derived as an effective theory in the
framework of supersymmetric  Grand 
Unified Theories (GUTs)~\cite{gut}, involving not
only the strong and electroweak interactions but gravity as well. 
A strong indication for the realization of this physical picture in
nature is the excellent agreement between the value of the weak
mixing angle $\sin^2\theta_W$ predicted by supersymmetric
gauge coupling unification  and the measured value
% one measured at the LEP1 experiment
~\cite{unif}-\cite{Nir2}. 
In addition to the unification of gauge couplings,  the unification of 
bottom and tau Yukawa couplings appears naturally in most  minimal 
supersymmetric GUTs and it determines the value of the  
top Yukawa coupling at low energies, 
providing an explanation for the heaviness of the 
top quark mass~\cite{CPW,Ramond}-\cite{BCPW}.
In the small to moderate $\tan \beta$ regime ($\tan \beta = 
v_2/v_1$, the ratio of the two Higgs vacuum expectation values),   
the condition of bottom-tau Yukawa unification  implies 
 a strong attraction of the top quark mass to 
its infrared fixed point,
yielding   a strong correlation between the top quark mass and $\tan \beta$.
The infrared fixed point solution has interesting implications for the 
Higgs, stop and chargino sectors of the theory~\cite{COPW}-\cite{ift}.
In the large $\tan\beta$ regime,  the phenomenology is more complex, mainly
because of  important supersymmetric threshold corrections to the 
bottom mass~\cite{Hall,wefour}. 
Of special interest in the large $\tan \beta$ region are 
theories with a richer symmetry structure, where all three
Yukawa couplings of the third generation unify, yielding a prediction for
both $M_t$ and $\tan \beta$~\cite{largetb}.

Minimal gauge coupling unification at a scale $M_{GUT}$
 of the order of 10$^{16}$ GeV 
 provides a prediction of one low energy gauge coupling as a function 
of the other two. One can then predict  $\sin^2\theta_W (M_Z)$
or, equivalently, consider as inputs the experimental
values of $\alpha_{em}(M_Z)$ and
 $\sin^2\theta_W(M_Z)$, and determine the value of the strong gauge coupling 
$\alpha_s(M_Z)$ to check if it is within
the experimentally accepted range, say 0.11--0.13. The experimental
prediction for  $\sin^2\theta_W(M_Z)$ in the modified $\overline{MS}$ scheme,
 in the limit of sufficiently heavy supersymmetric particles, can be  given as
a function of the electroweak parameters $G_F$, $M_Z$, $\alpha_{em}$ and 
the top quark mass;  it also depends to a lesser extent on the Higgs 
mass~\cite{LP,CPP}:
\bea
 \sin^2\theta_W(M_Z) &=&  0.23166 + 5.4 \times 10^{-6} (m_h - 100)           
  -   2.4 \times 10^{-8} (m_h -100)^2 
\nonumber \\
         &  & -  3.03 \times 10^{-5} (M_t - 165)
   -  8.4 \times 10^{-8}  (M_t - 165)^2 \pm 0.0003.
\label{eq:1}
\eea
Performing  the running of the gauge couplings
from the GUT scale down to $M_Z$ by  considering the  beta functions 
up to two-loops, one has also 
 to include properly the one--loop supersymmetric threshold 
corrections associated with the decoupling of the supersymmetric particles
at intermediate scales between $M_{GUT}$ and $M_Z$. 
For each  gauge coupling, the one-loop threshold correction 
to $1/ \alpha_i(M_Z)$
reads
\be
\frac{1}{\alpha_i^{thr.}} =  \sum_{\eta, M_{\eta} > M_Z} \frac{ b_i^{\eta}}
{2 \pi}  \ln \left(\frac{M_\eta}{M_Z} \right),
\ee
where the summation is over all sparticles and heavy Higgs bosons  with masses
$M_{\eta}$ larger than $M_Z$, and $ b_i^{\eta}$ is the 
contribution of each sparticle and  heavy Higgs to the one--loop 
beta function coefficient 
of the gauge coupling $\alpha_i$.
A detailed analysis shows that the above effect of supersymmetric thresholds
can be described in terms of one single scale $T_{SUSY}$~\cite {LP} which, 
considering different characteristic mass scales for
squarks ($m_{\tilde{q}}$),
 gluinos ($m_{\tilde{g}}$), sleptons ($m_{\tilde{l}}$),
 electroweak gauginos ($m_{\tilde{W}}$), 
Higgsinos ($m_{\tilde{H}}$) and the heavy
Higgs doublet ($m_H$), can be given as~\cite{CPW}
\be
T_{SUSY} = m_{\tilde{H}} \left( \frac{m_{\tilde{W}}}{m_{\tilde{g}}}
    \right)^{\frac{28}{19}} 
\left[ \left( \frac{m_{H}}{m_{\tilde{H}}} \right)^{\frac{3}{19}}
\left( \frac{m_{\tilde{W}}}{m_{\tilde{H}}} \right)^{\frac{4}{19}}
\left( \frac{m_{\tilde{l}}}{m_{\tilde{q}}} 
\right)^{\frac{3}{19}}
\right]
\label{eq:TSUSY}
\ee
The above relation holds whenever all the particles involved have masses  above
$M_Z$. If a mass is below $M_Z$ it should be replaced by $M_Z$ in the
computation of the threshold corrections to $\alpha_s(M_Z)$.
The value of the strong gauge coupling at low energies is then determined
as a function of 
 $\sin^2\theta_W$, $\alpha_{em}$ and $T_{SUSY}$. It 
follows that 
\be 
\frac{1}{\alpha_s(M_Z)} = \frac{1}{\alpha_s^{SUSY}(M_Z)} + \frac{19}{28 \pi} 
                             \ln \left(\frac{T_{SUSY}}{M_Z} \right),
\ee
where $ \alpha_s^{SUSY}(M_Z)$ would be the  value of the strong gauge coupling 
at $M_Z$ if the theory were exactly supersymmetric down to the scale $M_Z$.

Observe that  $T_{SUSY}$, Eq.~(\ref{eq:TSUSY}),
has only a slight dependence on the squark, slepton and heavy Higgs masses
and a very strong dependence on the overall Higgsino mass as well as on the 
masses of the gauginos associated with the electroweak and strong interactions.
In models with universal gaugino masses at the grand unification scale
it follows that, $T_{SUSY}
 \simeq  m_{\tilde{H}} \left( \alpha_2(M_Z)/ \alpha_s(M_Z)
\right)^{3/2} \simeq |\mu|/6$,
where $\mu$ characterizes the Higgsino mass in the case of negligible
mixing in the neutralino/chargino sector. Hence, if all supersymmetric masses
are $\simlt$ 1 TeV, the effective supersymmetric scale 
$T_{SUSY}$ is $\simlt$ the weak scale.

 The unification condition implies  the following numerical
 correlation~\cite{LP,Nir2,BCPW},
\be
 \sin^2\theta_W(M_Z) \simeq 0.2326 - 0.25 \left(\alpha_s(M_Z) - 0.123 \right)
\pm 0.0020
\label{eq:5}
\ee
where the central value corresponds to an effective supersymmetric threshold 
scale $T_{SUSY}= M_Z$ and the error is the estimated uncertainty in the 
prediction arising from a variation in $T_{SUSY}$ from 15 GeV to 1 TeV. 
Therefore, Eqs.~(\ref{eq:1}) and (\ref{eq:5}) imply that  
 the predictions from minimal gauge coupling unification agree with
the experimental data provided that
\bea
\alpha_s(M_Z) & = & 0.1268 + 1.21 \times 10^{-4} \left(M_t - 165 \right) 
   + 3.36 \times 10^{-7}  \left(M_t - 165 \right)^2 
\nonumber \\ 
 & & -  2.16 \times 10^{-5} \left(m_h - 100 \right) 
  + 9.6 \times 10^{-8}  \left(m_h - 100 \right)^2 \pm 0.009.
\eea
For values of the  top quark mass within the present experimental range,
 the above $M_t$--$\alpha_s$ correlation translates 
into acceptable  predictions for   $ \alpha_s(M_Z)$, 
which  have an important dependence
 on the range of the supersymmetric spectrum. Table 1 illustrates these
 results, including the two--loop effects
 of top and bottom  Yukawa couplings in the 
running of $\alpha_s$.

\begin{table}\begin{center}\caption{Gauge coupling unification 
predictions for $\alpha_s(M_Z)$, for 
given values of $\sin^2\theta_W$ (correlated with $M_t$),  $m_h = M_Z$
 and $T_{SUSY}$= 1 TeV $(M_Z)$}
\vspace{0.5cm}
\begin{tabular}{c|c|c} \hline\hline
& &  \\
$M_t$[GeV] & $\sin^2\theta_W(M_Z)$ & $\alpha_s(M_Z)$ \\ \hline
& &  \\
150 & 0.2321   & 0.116  (0.125) \\
& &  \\
170 & 0.2315   & 0.118   (0.127) \\
& &  \\
195 & 0.2306 &  0.122   (0.131)\\ \hline\hline
\end{tabular}
\end{center}
\end{table}

The predicted value of $\alpha_s(M_Z)$ from unification  may be further
modified if some sparticle masses are ${\cal{O}}(M_Z)$.
 Indeed, not only the leading--log
 contributions but the full one--loop threshold contributions from
SUSY loops should be included when extracting the couplings from
the data~\cite{all}. 
The main additional
effects come from light sfermions and  are given by \cite{all}-\cite{Nir2}
\begin{equation}
\frac{\delta \sin^2 \theta_W}{\sin^2 \theta_W}
\simeq \frac{\cos^2 \theta_W}{\sin^2 \theta_W - \cos^2 \theta_W}
\left( \frac{\delta \alpha_{em}}{\alpha_{em}} +
\frac{\Pi_{WW}(0)}{M_W^2} - \frac{\Pi_{ZZ}(M_Z^2)}{M_Z^2} 
\right),
\label{eq:delatsin}
\end{equation}
where 
%${\rm s}^2 = \sin^2 \theta_W$, ${\rm c}^2 = \cos^2 \theta_W$ and
$\Pi_{ij}$ are the vacuum polarization contributions to the gauge 
bosons. (The
standard model contributions and the leading--log contributions 
of sparticles heavier than the $Z$ boson are not included in 
Eq.~(\ref{eq:delatsin})  since they were already taken into
account previously.)
 Observe that, apart from a correction to the
$Z$ mass and a possible small correction to $\alpha_{em}$, 
the above expression is proportional to the parameter 
$\Delta \rho(0) = \Pi_{WW}(0)/M_W^2 - \Pi_{ZZ}(0)/M_Z^2$. In the MSSM
 it follows that
$(\Delta \rho(0))^{SUSY} \geq 0$ and,  after considering the 
additional terms in Eq.~(\ref{eq:delatsin}), one still obtains
 $\delta \sin^2\theta_W \leq 0$ in most of the
parameter space. This translates into an increase, 
with respect to the  results from table 1, in the
values of $\alpha_s(M_Z)$  predicted from supersymmetric grand unification,
 which may be important when
the supersymmetric spectrum is sufficiently light. 
Larger values of $\alpha_s(M_Z)$ may however be in conflict with the data.
(One should also keep in mind that large
corrections to $\Delta \rho (0)$ are disfavoured by  present experimental
data, particularly for large values of the top quark mass
 $M_t \geq 175$ GeV.)
High energy threshold corrections may be helpful in
moderately lowering the $\alpha_s(M_Z)$ prediction. In fact, in the above
I have considered the minimal gauge coupling unification scenario,
that is, neglecting all possible GUT/$M_{Planck}$ 
scale threshold corrections in the
running of the couplings. Perturbations to the unification relations
 are in general strongly model  
dependent, and they must be appropriately computed once a given high 
energy model is specified~\cite{GUTthr}.

For the present experimental range of values
 for the top quark mass~\cite{tevatron}, 
$M_t = 180 \pm 12$  GeV,
the condition of bottom-tau  Yukawa coupling 
unification implies either low values of
$\tan \beta$, $ 1\leq \tan \beta \leq 3$, or very large values 
of $\tan \beta = {\cal{O}} (m_t/m_b)$~\cite{Ramond}-\cite{BCPW}.
Most interesting is the fact that to achieve b-$\tau$ unification,
 $h_b(M_{GUT}) =  h_{\tau}(M_{GUT})$, large values
 of the top Yukawa coupling, $h_t$, at $M_{GUT}$
are necessary in order to compensate for the effects of the strong 
interaction renormalization in the running of the bottom Yukawa coupling. 
These large 
values of $h_t^2(M_{GUT})/4 \pi \simeq$ 0.1--1 are exactly those that 
ensure the attraction towards the infrared (IR) 
fixed point solution of the top
quark mass~\cite{CPW,Yukawa,LP1}. 
In fact, the strength of the strong gauge coupling as well as 
the experimentally allowed range of 
values of the bottom mass play a decisive role
in this behaviour~\cite{BCPW}. 
For values of $\alpha_s(M_Z) \geq 0.115$, 
which are those preferred by the condition of minimal 
gauge coupling unification,
the strong gauge coupling is sufficiently strong to demand a large value
 of the top Yukawa coupling to contravene its renormalization effect on the
 bottom mass\footnote{ Smaller 
values of $\alpha_s(M_Z)$, as could be obtained 
in the presence of large high energy threshold corrections in the running
of the couplings, would change this picture.}.
 Also the fact that the pole bottom mass $M_b$ is in the 
4.6--5.2 GeV range is extremely important~\cite{BCPW,KKRW}.
 A larger $M_b$, for instance, will
be associated to a larger bottom Yukawa coupling allowing unification
for the same values of $\alpha_s$ with a weaker top Yukawa coupling, relaxing
hence  the
infrared fixed point attraction. In summary, in the low $\tan \beta$ case
one obtains that
for the presently allowed values of the electroweak parameters and the
 bottom mass,  b-$\tau$ unification implies that the top quark mass
must be within ten per cent of its infrared fixed point values. A mild 
relaxation of exact unification (0.9 $ \leq 
 h_b/h_{\tau}|_{M_{GUT}} \leq$ 1)
still preserves this feature, especially for 
 $M_b \leq 4.95$ GeV \cite{EPS,NIRlast}. 
In the large $\tan \beta$ region, instead,
one has $h_b = {\cal{O}}(h_t)$ and, hence,
within the context of b--$\tau$ unification,
 the infrared fixed point attraction
is much weaker than for low values of $\tan \beta$.

As mentioned above, the fixed point solution, $h_t=h_t^{IR}$,
is obtained for large values of the top Yukawa 
coupling  at the grand unification scale, which still remain in the
 perturbative regime.  For $M_{GUT} \simeq 10^{16}$ GeV one obtains at low
energies $(h_t^{IR})^2/ 4 \pi \simeq
(8/9) \alpha_s(M_Z)$, and the running top quark mass tends to its
infrared fixed point value $m_t^{IR} = h_t^{IR} v \sin \beta$, with 
$v \simeq 174$ GeV. Hence, relating the running top quark mass $m_t$
 with the pole
top quark mass $M_t$ by considering the appropriate QCD corrections, for 
$\alpha_s(M_Z)$ in the range 0.11--0.13 and small or moderate
$\tan \beta$ values, one has~\cite{CW}\footnote{In Eq. \ref{eq:MtIR},  
low energy SUSY threshold corrections, which
 may have a mild incidence in the definition of the infrared fixed point
solution of the top quark mass, has been neglected \cite{NIRlast,NIRref}.},
\bea 
m_t^{IR}(M_t) &\simeq & 196 {\rm GeV} \;
 [1+ 2\left(\alpha_s(M_Z)- 0.12 \right)] \; \sin \beta \; ;
\nonumber \\
M_t^{IR} & =  &m_t^{IR}(M_t) \;
\left[1 + \frac{4 \alpha_s(M_Z)}{3 \pi} + {\cal{O}}(\alpha_s^2) \right]. 
\label{eq:MtIR}
\eea

The infrared fixed point structure also plays a decisive role in the 
evolution of the fundamental mass parameters of the theory, which 
has important implications on the Higgs 
and supersymmetric particle spectra~\cite{COPW,BABE}-\cite{ift}.
The  $M_t$--$\tan\beta$ relation, Eq.~(\ref{eq:MtIR}), associates to 
each value of $M_t$ the lowest possible value of $\tan \beta$ consistent
 with the validity of perturbation theory up to scales of order $M_{GUT}$.
In addition, the infrared fixed point solution, $h_t 
\rightarrow h_t^{IR}$, induces an infrared 
fixed point value of the trilinear coupling 
$A_t$~\cite{COPW,KPRZ}\footnote {Nontrivial fixed points can also be present 
 at high energies, leading to predictions for some supersymmetry 
breaking parameters at the GUT scale \cite{Ross}.}.
 One has 
$A_t \simeq A_0 [1 -
 h_t^2/(h_t^{IR})^2]
 - M_{1/2} [4 - 2 h_t^2/(h_t^{IR})^2]$,
  with $A_0 = A_t(M_{GUT})$ and $M_{1/2}$
the common gaugino mass at $M_{GUT}$. Hence, $A_t^{IR} \simeq - 2 M_{1/2}$
and this results in a very weak
dependence of the whole spectrum on the parameter $A_0$. In addition,
the combination of soft 
supersymmetry breaking mass parameters $M_{QU}^2 = m_Q^2 +  m_U^2 + 
m_{H_2}^2 $
 has an IR fixed point behaviour as well: $M_{QU}^{IR}
 \simeq \sqrt{6.5} M_{1/2}$.
Moreover, the condition of a proper electroweak symmetry breaking determines 
the supersymmetric mass parameter $\mu$ at the IR in terms of $\tan \beta$,
$M_{1/2}$ and the common scalar mass $m_0$ at the unification scale. Therefore,
in the case of universal mass parameters, 
$M_{1/2}$ and $m_0$ at $M_{GUT}$, for
 any given value of the top quark mass at the IR fixed point,
 the whole spectrum is basically 
determined in terms of only these two high--energy parameters.
 The effect of non-universal $m_0$ on the spectrum
 is, however, very interesting to study as well~\cite{CW,nonuniv}.

The most
 interesting consequence of the 
IR fixed point $M_t$--$\tan\beta$ relation is associated with
 the lightest CP-even  Higgs mass predictions in the MSSM~\cite{COPW,BABE}. 
Indeed, for $\tan\beta$ larger than 1, the lowest tree level value of the 
lightest 
Higgs mass, $m_h$,
is obtained  at the lowest value of $\tan\beta$. 
Hence, since in any theory
consistent with perturbative unification the IR fixed point 
solution assignes the lowest possible $\tan \beta$ value to each $M_t$, 
by the same token, the IR fixed point solution is
associated with the lowest value of 
the tree level mass of the lightest Higgs boson
consistent with the theory. Even after the inclusion of radiative corrections,
the upper bound on  the lightest Higgs mass
 is considerably reduced at the
fixed point solution: considering  $M_t$ = 160, 170, 180, 190 GeV, 
for a characteristic supersymmetric
 mass scale of 1 TeV, the upper limit  yields
 $m_h^{IR} \leq$  80, 90, 110, 127  GeV~\cite{CEQW},
which is considerably smaller than the upper bound one would obtain in the 
general MSSM framework, $m_h \leq$ 120, 125, 132, 140 GeV~\cite{CEQW},
 respectively.
This is of particular interest for  experimental searches. In fact,
 for $M_t \simlt 175$ GeV, if the infrared  fixed point top quark 
mass solution is
realized in nature, the lightest CP-even Higgs mass must be within the
reach of LEP2 for $\sqrt{s} =$ 192 GeV~\cite{interim}. 

The infrared fixed point solution is also  experimentally appealing 
in relation to the sparticle spectrum.
Considering the behaviour of the mass parameters of the theory it
 follows that light charginos, $m_{\tilde{\chi}^+} \simlt $ 80 GeV,
 and light right--handed stops, $m_{\tilde{t}} \simlt$
120 GeV, may be 
present in the theory. This is of particular interest for direct experimental
searches, and it can  induce important positive corrections in 
$R_b = \Gamma (Z \rightarrow b\bar{b})/ \Gamma (Z \rightarrow hadrons)$, 
which 
can ameliorate the present discrepancy between the experimentally
 measured value and the SM prediction for this
quantity \footnote {Very large corrections to $R_b$ imply, however,  that 
the value of $\alpha_s(M_Z)$
derived from experiments is shifted towards lower values, enhancing the 
necessity for large high-energy threshold corrections to 
the running of the gauge couplings in the framework of
 unification.}~\cite{CW,Gordy,Rb}.
%The best value which can be obtained in the IR fixed point framework, without
%destroying the predictions for other measurable quantities, is $R_b \simeq$
%0.218--0.219.
 To obtain sizeable corrections to  $R_b$, the lightest chargino 
must be mainly Higgsino like.  Within  the IR fixed point solution this
 necessarily demands  to relax 
the universality condition for the scalar soft supersymmetry breaking masses
at the unification scale~\cite{CW}.

In the context of b--$\tau$ Yukawa coupling unification it is clear  that,
possible large  radiative corrections to the bottom
 mass are crucial in determining
 the top quark mass and  $\tan \beta$ predictions.
In general, one assumes that the top and bottom quarks couple each to only one
of the Higgs doublets and hence $m_t (M_t) =  h_t (M_t) v_2$
and $m_b (M_t) =  h_b (M_t) v_1$,
with $v_i$ the vacuum expectation value of the Higgs $H_i$. 
 However, a coupling of the bottom (top) quark to the 
neutral component of the Higgs $H_{2(1)}$ may
be generated at the one--loop level, and,
for large values of $\tan \beta \simgt 40$, since $v_2 \gg v_1$, large    
 corrections to the bottom mass may be present~\cite{Hall,wefour,CDWR}, 
\bea
m_b &=& h_b \; v_1 + \Delta h_b \; v_2 \equiv \tilde{m_b} (1 + K \tan \beta).
\eea
$\Delta m_b = K \tan \beta$
receives contributions from stop--chargino and sbottom--gluino loops,
the latter being the dominant ones.
The magnitude of  $\Delta m_b$ is strongly dependent on the supersymmetric
spectrum and its sign is generally 
governed by the overall sign of $\mu \times m_{\tilde{g}}$.
\bea
\Delta m_b &  =&  \mu \; m_{\tilde{g}} \;\tan \beta \;\;
 \left[ \frac{2 \alpha_s}{3 \pi}
I_1 ( m^2_{\tilde{b}_1}, m^2_{\tilde{b}_2},  m^2_{\tilde{g}})
 +   \frac{A_t}{ m_{\tilde{g}}} \frac{h_t^2}{(4 \pi)^2}
I_2 ( m^2_{\tilde{t}_1}, m^2_{\tilde{t}_2}, \mu^2) \right],
\label{eq:Deltamb}
\eea
where $ m_{\tilde{q}_i}$ are the squark mass eigenstates and the integral
factor is $I_i = K_i/a_{max}$,
 with $a_{max}$ the maximum of the squared masses and
$K_i$ = 0.5--0.9 depending on the mass splitting. Using the relation
$ m_{\tilde{g}} \simeq $ 2.6--2.8 $M_{1/2}$ and the fact that from the 
renormalization group  equations $A_t$ is in general of  opposite  
sign and of ${\cal{O}} (M_{1/2})$, it follows that there is a partial 
cancellation between the two terms in Eq.~(\ref{eq:Deltamb}). Although 
important, such partial cancellation 
is by far not sufficient to render the bottom mass corrections small.
Hence, in the large $\tan \beta$ region, the bottom mass corrections
need to be appropriately computed and their final effect on the predictions
from b--$\tau$ Yukawa coupling unification will depend on the particular
 supersymmetric spectrum under consideration.

Large values of $\tan\beta$ are also interesting from the
point of view of precision measurements \cite{Gordy,Rb,Joan}.
Indeed, the fit to the measured
value of $R_b$ can also be improved for large values of
$\tan\beta \simeq m_t/m_b$, particularly if the CP-odd Higgs mass is below
70 GeV.
This is due to the  large one-loop positive corrections 
associated with the neutral CP-odd Higgs scalar sector of the 
theory, which become important when the supersymmetric
bottom quark Yukawa coupling is enhanced. 
  Experimentally, low values of the
CP-odd Higgs mass, $m_A \simlt m_Z$, are clearly very interesting 
 from the point
of view of  Higgs searches, since they imply that both the lightest
CP-even and the CP-odd Higgs masses will be within the reach of LEP2, 
$m_h \simeq m_A$.
The charged Higgs mass is approximately determined through the 
CP-odd Higgs mass value, $m_{H^{\pm}}^2 \simeq m_A^2 + M_W^2$, as well, and 
hence, strong constraints can be obtained
on the Higgs spectrum  by considering the
charged Higgs contributions to the  branching ratio 
 ${\rm BR}(b \rightarrow s \gamma)$.
Even taking into  account in a very conservative way the QCD uncertainties
associated with the
${\rm BR}(b \rightarrow s \gamma)$ (assuming 40$\%$ QCD uncertainties), 
for  $m_{H^{\pm}}
\simlt 130$ GeV the  
$b \rightarrow s \gamma$ decay rate becomes larger than the
presently allowed experimental values~\cite{bsgexp}, unless 
supersymmetric particle contributions suppress the 
charged Higgs enhancement of the decay rate.
The most important supersymmetric
contribution to this rare bottom decay mode
comes from the chargino-stop one-loop diagram~\cite{bsga}. 
The chargino contribution to the $b \rightarrow s \gamma$
decay amplitude depends on the soft supersymmetry breaking
mass parameter $A_t$ and on the supersymmetric mass parameter
$\mu$, and for very large values of $\tan\beta$, it is
given by~\cite{Diaz,BOP},
\be
A_{\tilde{\chi}^+} \simeq \frac{m_t^2}{m_{\tilde{t}}^2}
\frac{A_t \mu}{m_{\tilde{t}}^2} \tan\beta \; G\left(
\frac{m_{\tilde{t}}^2}{\mu^2}\right),
\ee
where $G(x)$ is a function that takes values of order 1
when the characteristic stop mass $m_{\tilde{t}}$
is of order $\mu$ and grows for lower values
of $\mu$. One can show that, for positive (negative)
values of $A_t \times \mu$ the chargino contributions are of
the same (opposite) sign as the charged Higgs ones~\cite{wefour}. 
Hence, to partially cancel the light charged Higgs contributions rendering
the  $b \rightarrow s \gamma$
decay rate acceptable, negative values for $A_t \times \mu$ are required.
As  follows from Eq.~(\ref{eq:Deltamb}) and the discussion below,
 this requirement
has direct implications on the corrections to the bottom mass
and,  after a detailed analysis, one concludes that it puts
strong constraints on models with Yukawa 
coupling unification~\cite{wefour,BOP}.
Performing a $\chi^2$ fit to precision data, it follows that a light
Higgs mass regime  with unification of
 the three Yukawa couplings of the third generation  at $M_{GUT}$ is possible 
but, the soft supersymmetry
breaking parameters at high energies need to be highly  non--universal.
For  moderate values of the soft supersymmetry  breaking parameters  and 
$\alpha_s(M_Z) \simeq 0.125,0.120,0.115$,
the 
top quark mass is in the range $M_t \simgt 180,170,160$ GeV,
 respectively~\cite{wefour,we5}. 

In this talk I have considered the MSSM with three generations of quark 
and lepton superfields. The framework  of unification of couplings  can also be
explored in models with four generations. Different fourth-generation
low energy supersymmetric scenarios with
a heavy \cite{Gunion} or a light 
($M_t \simeq M_W$)\cite{Howienos} top quark have been studied.
In these cases one obtains
interesting predictions for some of the third- and fourth-generation 
particle masses, which can be tested at the 
 Tevatron  or at the next run of the  LEP collider.\\
~\\
{\bf Acknowledgements:}
I would like to thank W. Bardeen, P. Chankowski, S. Dimopoulos, J.-R. Espinosa,
M. Olechowski, S. Pokorski,
M. Quiros, S. Raby and C. Wagner,
for many fruitful and enjoyable collaborations, 
in which we studied many of the  subjects discussed in this talk. \\
~\\

\end{document}